\documentclass[graybox]{svmult}

\usepackage{type1cm}        
                        
\usepackage{wrapfig}

\usepackage{makeidx}         
\usepackage{graphicx}        
\usepackage{multicol}        
\usepackage[bottom]{footmisc}

\usepackage{standalone}
\usepackage{floatrow}
\usepackage{newtxtext}       %
\usepackage[varvw]{newtxmath}       
\usepackage{pgfplots}
\usepackage{pgfplotstable}
\usepackage{tikz}
\usepackage[capitalize]{cleveref}
\usepackage{physics}
\usepackage{amsmath}
\usepackage{mathtools}
\usepackage[font=small,skip=0pt]{caption}
\usepackage{globalplotconfig}
\usepackage{customcolours}
\usepackage{customcommands}
\pgfplotsset{compat=newest}
\usetikzlibrary{pgfplots.groupplots}
\usepackage[normalem]{ulem}

\usetikzlibrary{decorations,calligraphy,positioning,calc,math}
\usetikzlibrary{decorations,calligraphy, positioning, calc, math, patterns,
  decorations,
  decorations.markings,
}

\makeindex             

\begin{document}


\title*{Multi-target search in bounded and heterogeneous environments: a lattice random walk perspective}
\titlerunning{Multi-target search in discrete space and time} 
\author{Luca Giuggioli, Seeralan Sarvaharman, Debraj Das, Daniel Marris and Toby Kay}
\institute{Luca Giuggioli, \at University of Bristol, School of Engineering Mathematics and Technology, \email{luca.giuggioli@bristol.ac.uk}
\and Seeralan Sarvaharman \at University of Bristol, School of Engineering Mathematics and Technology, \email{seeralan.sarvaharman@mpinat.mpg.de.ac.uk}
\and Debraj Das \at Quantitative Life Sciences, ICTP -- The Abdus Salam International Centre for Theoretical Physics, Trieste 34151, Italy, \email{ddas1@ictp.it}
\and Daniel Marris \at University of Bristol, School of Engineering Mathematics and Technology, \email{daniel.marris@bristol.ac.uk}
\and Toby Kay \at University of Bristol, School of Engineering Mathematics and Technology, \email{t.kay@imperial.ac.uk}}

%
%
\maketitle


\abstract{For more than a century lattice random walks have been employed ubiquitously, both as a theoretical laboratory to develop intuition about more complex stochastic processes and as a tool to interpret a vast array of empirical observations. Recent advances in lattice random walk theory in bounded and heterogeneous environments have opened up opportunities to cope with the finely resolved spatio-temporal nature of modern movement data. We review such advances and their formalisms to represent analytically the walker spatio-temporal dynamics in arbitrary dimensions and geometries. As new findings, we derive the exact spatio-temporal representation of biased walks in a periodic hexagon, we use the discrete Feynman-Kac equation to describe a walker's interaction with a radiation boundary, and we unearth a disorder indifference phenomenon. To demonstrate the power of the formalism we uncover the appearance of multiple first-passage peaks with biased walkers in a periodic hexagon, we display the dependence of the first-transmission probability on the proximity transfer efficiency between two resetting walkers in a one-dimensional periodic lattice, we present an example of spatial disorder in a two-dimensional square lattice that strongly affects the splitting probabilities to either of two targets, and we study the first-reaction dynamics to a single lattice site in an unbounded one-dimensional lattice.}

\section{Introduction}

The dynamics of reaction-diffusion processes in heterogeneous space is a topic with vast applicability from transport processes in physical and chemical systems \cite{ben-avrahamhavlinbook2004} to animal ecology and disease spread \cite{kenkregiuggiolibook2021}. While the literature encompasses both continuous and discrete variable descriptions, we focus here on the latter providing an overview of the discrete space-time formalism. Such formalism is the lattice random walk (LRW) whose history (see e.g. \cite{todhunter1865} for an historical account of probability theory) dates back to the early works by de Moivre \cite{demoivre1756} and Laplace \cite{laplace1820} and to seminal contributions, during the last century, by Smoluchowski \cite{smoluchowski1906b}, Polya \cite{polya1921}, Chandrasekhar \cite{chandrasekhar1943}, Kac \cite{kac1947}, Feller \cite{feller1950}, Erd\"{o}s \cite{dvoretzkyerdos1951}, Sparre Andersen \cite{sparreandersen1954} and Spitzer \cite{spitzer1976}, with more recent monographs by Weiss \cite{weiss1994} and Hughes \cite{hughesbook1995}.

The studies of LRW dynamics in spatially heterogeneous space can be traced back to the work by Montroll in the 60's to study trapping reactions in photosynthetic units \cite{montroll1964random,montrollweiss1965,montroll1969}. Since then a large literature on the transport properties of random walks both in continuous and discrete time emerged \cite{kenkre1980,denhollanderkasteleyn1982,rubinweiss1982,kenkre1982,derrida1983,szaboetal1984,hauskehr1987,murthykehr1989,hughesbook1996,cordesetal2001}, but it mainly focused on translationally invariant one-dimensional systems. Interest for theoretical studies in non-periodic bounded domains have surfaced more recently (see e.g. \cite{redner2001,benichouvoituriez2014,metzleretal2014b}). Part of the more recent surge of interest stems from the technological progress in various types of sensors and instruments that allow to record with unprecedented resolution and quality the spatio-temporal dynamics of physical particles and biological organisms as well as of the environment where motion occurs.

These advances demand the development of models that account in an explicit manner, rather than simply averaging over the disorder, for how random movement is affected by the presence of spatial heterogeneities. More broadly, there is a need for models capable of describing how the underlying 
space affect motion. Very recently a series of exact analytic studies on the dynamics of (Markov) LRWs in arbitrary dimensions \cite{giuggioli2020,sarvaharmangiuggioli2020,dasgiuggioli2022,marris2023exact,giuggiolisarvaharman2022,sarvaharmangiuggioli2023} have opened up this possibility. Here we review this recent literature, what knowledge gaps it has filled and what it has accomplished. We also present some new findings and we point to open problems where LRW can provide some useful contribution.

\subsection{The missing links}
\label{sec:mis_link}
\newcommand{\basis}[1][]{\ensuremath{\mvec{e}_{{#1}}}}
While mathematical convenience has often favoured studies of diffusive transport processes with continuous space description---no need for spatial discretisation---the benefit of discrete space models is that the dynamics of many relevant quantities can now be found in closed form. This benefit is especially relevant in dimensions greater than one where finding the spatio-temporal dependence of the diffusion equation, except in very symmetric geometries, requires the numerical solution of complicated boundary value problems. In addition, when time is discrete, that is when dealing with LRWs, the inversion to time of closed form expressions of generating functions is computationally more convenient when compared to a numerical Laplace inversion to continuous time.

Given these advantages, one may wonder why LRWs have not been employed more extensively to study first-passage processes in heterogeneous space. The answer to this question lies in the multiple technical challenges, our missing links, that needed to be overcome, starting from the analytic knowledge of the LRW spatio-temporal dependence of the occupation probability $P_{\nvec_0}(\nvec,t)$, where $\nvec$ represents the lattice coordinate, $\nvec_0$ the localised initial lattice and $t$ the discrete time variable. In unbounded hypercubic lattices the exact solution of the LRW Master equation, the so-called propagator or Green's function, for nearest-neighbour jumps can be easily found for each Fourier component $\left(\widehat{f}(k_i)=\sum_{n_i= - \infty}^{+\infty} \mathrm{e}^{ik_in_i}f(n_i)\right)$ and as a generating function $\left(\ztrans{f}(z) = \sum_{t = 0}^{\infty} f(t)z^t\right)$, namely in the form $\widetilde{\widehat{P}}_{\nvec_0}(\boldsymbol{k},z)$. The same could not be said, instead, about finite lattices of arbitrary dimensions for which exact propagators have remained unknown for over a century \cite{giuggioli2020}, except for periodic domains \cite{montroll1956}. In Sec. \ref{sec:euclid} we present the development that led to the exact spatio-temporal dependence of $P_{\nvec_0}(\nvec,t)$ in hypercubic lattices with arbitrary boundaries, i.e. absorbing, periodic, reflecting and mixed absorbing/reflecting.

Knowledge of the propagators has allowed us to study in closed form the LRW dynamics in the presence of spatial disorder \cite{dasgiuggioli2023,sarvaharmangiuggioli2023}. For the latter, and to make a distinction with the dynamics in the presence of reactive sites, we have used the name inert spatial heterogeneities to indicate that the LRW occupation probability is conserved over time. Such heterogeneities consist e.g. of permeable and reflecting barriers, regions with different diffusivities, local biases as well as long-range connections. We show this advance in Sec. \ref{sec:inert}. A special type of inert heterogeneity, that affects the movement dynamics globally, occurs when a LRW jumps with some probability from anywhere on the lattice to a specific site. These dynamics represent the so-called resetting LRW in discrete space and discrete time, originally introduced in the literature for Brownian walks in ref. \cite{evansmajumdar2011}. The form of the resetting discrete Master equation, derived and solved in ref. \cite{dasgiuggioli2022}, is presented in Sec. \ref{sec:reset}. Use and extension of the above approaches to other geometries has also been possible, one such important endeavour has been the exact derivation of the
LRW dynamics in finite periodic and reflecting hexagonal and honeycomb domains in ref. \cite{marris2023exact}, and have remained unknown for over a century \cite{giuggioli2020}. In Sec. \ref{sec:hex} we report on these latter findings, but we also extend them and find  the propagator when an hexagonal LRW is subject to global biases.

In Sec. \ref{sec:fpp} we present the mathematical formalism \cite{giuggiolisarvaharman2022} that allows us to determine the splitting probabilities when the environment contains multiple partially or fully absorbing targets. As studies of radiation boundary conditions for LRWs have been very limited, we have dedicated Sec. \ref{sec:rad} to introduce the formalism and show the link to the discrete Feynman-Kac equation \cite{kac1949,kac1951,csaki1993}, which allows to analyse functionals of random processes. In Sec. \ref{sec:fpfeft} we cover four topics and we do so by applying the formal theories presented in the preceding sections. We study the dynamics of the first-passage probability with a single target on a periodic hexagonal geometry, the splitting probabilities to two targets in a heterogeneous 2D reflecting square lattice, the transmission probability between two resetting LRWs in a 1D periodic domain, and the first-reaction probability in a 1D unbounded domain. The analytical nature of the formalism is exploited in discovering  an interesting disorder indifference phenomenon in Sec. \ref{sec:dis_indif}. While for a semi-bounded 1D domain a disorder indifference phenomenon has already been presented in the literature \cite{giuggiolisarvaharman2022}, here we show another example, which occurs in a 2D domain. Finally, in the concluding Sec. \ref{sec:concl}, we point to research areas where our formalism could be further exploited as well as to areas where LRWs can provide fruitful approaches.


\section{Dynamics in bounded hypercubic lattices}
\label{sec:euclid}
The first step in identifying the missing links
has been to devise a hierarchical procedure that exploits the analytic solution in a lower dimension to build the solution in a higher one \cite{giuggioli2020}. Such procedure, which involves a symmetrisation step in the presence of global biases \cite{sarvaharmangiuggioli2020}, has allowed to find the exact spatio-temporal solution of the nearest-neighbour LRW Master equation in hypercubic lattices of $d$ dimensions, namely
\begin{eqnarray}
  \label{eq:bias_nd_master}
    \proptt[t+1]{\nvec}{} =
      \Bigg(1 - \frac{1}{d}\sum_{i =1}^{d}q_i\Bigg) \proptt{\nvec}{} 
  &+&\sum_{i =1}^{d}\Bigg\{  \frac{q_i}{2d} (1 - g_i) \proptt{\nvec - \basis[i]}{} \Bigg.  \nonumber \\
   &+& \Bigg. \frac{q_i}{2d} (1 + g_i) \proptt{\nvec + \basis[i]}{} \Bigg\},
\end{eqnarray}
where $\basis[i]$ is the $i$-th basis vector and where, at each time step, $-1\leq g_i\leq 1$ represents the bias along the $i$-th axis ($g_i>0$ indicates a bias in the negative direction), and with $0 < q_i \leq 1$ for all $i$ yielding $0\leq 1-  d^{-1}\sum_{i=1}^dq_i<1$ the probability of not moving. Given that the  $q_i$'s are proportional to the diffusion constant along each direction in the continuous space-time limit of Eq. (\ref{eq:bias_nd_master}), we refer to them as the LRW diffusivities.

In bounded hypercubic lattices, once the types of domain boundary along each axis are specified, the propagator of Eq. (\ref{eq:bias_nd_master}) for the initial condition $P_{\mvec{n}_0}(\mvec{n},0) = \delta_{\mvec{n},\mvec{n}_0}$, with $\delta_{\mvec{a},\mvec{b}}$ representing the Kronecker delta fucntion for each axis, is given by \cite{giuggioli2020}
\begin{align}
\proptsymbol^{(\mvec{\gamma})}_{\mvec{n}_0}(\mvec{n}, t)\!=\!\sum_{k_1 = w^{(\gamma_1)}}^{W^{(\gamma_1)}}\!\cdots\!
  \sum_{k_d = w^{(\gamma_d)}}^{W^{(\gamma_d)}}\ \prod_{j=1}^{d} h_{k_j}^{(\gamma_j)}(n_{_j}, n_{0_j})
\ls 1 + \frac{s_{k_1}^{(\gamma_1)}}{d}\!+\!\cdots\!+\!\frac{s_{k_d}^{(\gamma_d)}}{d} \rs^t ,
\label{bias:eq:full_nd_prop}
\end{align}
where the superscript $\gamma$, accounting along each axis for domains that are periodic ($\gamma=p$), absorbing ($\gamma=a$), reflecting ($\gamma=r$) or a mixture of reflecting on one side and absorbing on the other ($\gamma=m$), specifies the spatial and temporal dependence.
\begin{figure}[!htbp]
\centering
\includegraphics{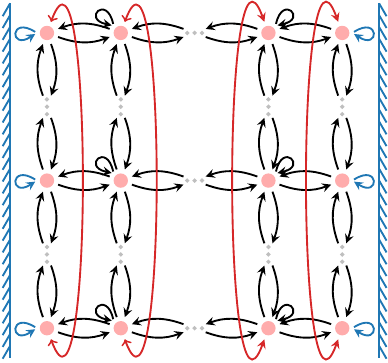} \hfill \includegraphics[width = 0.395\textwidth]{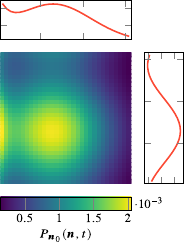}
\caption{Left panel: Schematic representation of the transition probabilities for a biased LRW on a 2D square lattice limited by reflecting and periodic boundaries along the horizontal and vertical axes, respectively. The black arrows away from the boundaries denote transition probabilities in the bulk with the following values: $q_1(1+g_1)/2$ for horizontal left steps, $q_1(1-g_1)/2$ for horizontal right steps, $q_2(1+g_2)/2$ for vertical steps down, $q_2(1-g_2)/2$ for vertical steps up, and $1-(q_1+q_2)/2$ for the slanted oval, representing the probability of staying. Two-headed red long arrows depict transition probabilities across the periodic boundary and therefore have the same values as the vertical black arrows with corresponding directions. The blue oval arrows at the left and right extremes denote the modified probabilities of staying because of the reflecting boundaries along the horizontal directions. The values of these arrows on the left and right extremes are given by $1+q_1g_1/2-q_2/2$ and $1-q_1g_1/2-q_2/2$, respectively.
Right panel: Corresponding occupation probability and its marginals (top and side panels) at time $t=200$ of the biased LRW with $q_1=q_2=0.8$ and $g_1=g_2=0.1$, starting at $\boldsymbol{n}_0=(21,21)$, in a square domain of size $N=31$ with reflecting and periodic boundaries along, respectively, the horizontal and vertical axis. With the definitions $f=(1-g)/(1+g)$ and $\sigma=(1-g^2)^{-1/2}$, the temporal dependence in Eq. (\ref{bias:eq:full_nd_prop}) is given by
$s^{(p)}_{k}=q\cos(2\pi k/N)+iqg\sin(2\pi k/N)-q$ for the periodic axis, while for the reflective axis we have $s^{(r)}_0=0$ or $s^{(r)}_{k}=q\cos(\pi k_i/N)/\sigma-q$ when $k>0$. For the spatial dependence we have $h^{(p)}_k(n,n_0)=\exp[2\pi i k(n-n_0)/N]/N$, while $h^{(r)}_0(n,n_0)=f^{n-1}(1-f)/(1-f^N)$ and $h^{(r)}_k(n,n_0)=N^{-1}f^{(n-n_0-1)/2}g(n)g(n_0)/[\sigma-\cos(\pi k/N)]$ with $g(m)=\sqrt{f}   \sin(\pi km/N)-\sin[\pi k(m-1)/N]$ when $k>0$. The finite series for the $k_i$ indices start with $\omega=0$ and terminates at $W=N-1=30$.}
\label{fig:biaslrw}
\end{figure}
While we refer the reader to refs. \cite{giuggioli2020,sarvaharmangiuggioli2020} for the explicit expressions for $h_{k}^{(\gamma)}(n,n_0)$, $s_{k}^{(\gamma)}$, $w^{(\gamma)}$, and $W^{(\gamma)}$, we plot in Fig.~\ref{fig:biaslrw} an illustrative example, that of a LRW in a square domain periodic and reflecting along two orthogonal directions, globally biased towards the bottom left corner. At intermediate times the occupation probability, $P_{\boldsymbol{n}_0}(\boldsymbol{n},t)$, displays some non-monotonicity in space along the horizontal axis due to reflection, while a non-zero probability appearing close to the top boundary indicates the presence of the periodic boundary conditions along the vertical axis. At step $t=200$ we observe the peak of the probability at coordinates (13,13), which is 16 lattice sites away from the starting location in the South West direction. To explain this observation we calculate the ensemble average, indicated with the symbol $\big\langle \big\rangle$, using the unbounded propagator and obtain the mean displacement $\big\langle n_i\big\rangle=
\sum_{\nvec} n_i P(\nvec, t)
=n_{0_i}-g_iq_it/2$, where $ i= 1$ and 2 denotes, respectively, the horizontal and vertical directions. To determine the time dependent mean position of the probability we simply subtract the initial position along each axis and we have $\big\langle(n_i-n_{0_i})\big\rangle=-g_iq_it/2$, which is exactly -8 along each axis for the parameters used in Fig. \ref{fig:biaslrw}. An analogous calculation for the mean square displacement (MSD) extracted from the unbounded propagator along each axis gives $\big\langle\big(n_i-\big\langle n_i\big\rangle\big)^2= q_it\left(1-g_i^2q_i/2\right)/2$, whose square root gives a distance of around 9  for the parameters of Fig. \ref{fig:biaslrw}. With the square domain of width 31 and the mode centred at $(13,13)$ at time $t=200$, it becomes clear why the effects of the boundaries start to be conspicuous at such time scale, and are particularly visible along the horizontal axis because of the probability accumulation due to the reflecting boundary.

Differently from the corresponding cases for random walks in continuous time as well as Brownian walks, one may notice that Eq. (\ref{bias:eq:full_nd_prop}) is not the product of 1D propagators, which ultimately has made identifying the exact spatio-temporal solution of the Master equation challenging. Another notable characteristic of Eq. (\ref{eq:bias_nd_master}) and its solution is that the dynamics include the chance at each time step of staying at a site with some arbitrary probability. In the mathematics literature a symmetric nearest-neighbour LRW that stays put with probability 1/2 and moves in all allowed directions with probability $1/(4d)$ is called a lazy random walk \cite{hildebrand2001,millerperes2012,kovac2023}. In our Eqs. (\ref{eq:bias_nd_master}) and (\ref{bias:eq:full_nd_prop}), we obtain such kind of transition probability by setting $g_i=0$ and $q_i=1/2$. This implies that our formalism generalises the lazy LRW, allowing to describe its occupation probability when the chance of not moving at each time step can be chosen arbitrarily along each direction, a practical convenient aspect since it avoids the so-called parity issues \cite{kac1947a} that affect the spatio-temporal dependence of the LRW occupation probability dynamics when $q_i=1$. From now on we refer more broadly to a lazy LRW whenever a walker at each time step has a non-zero chance of staying.

\subsection{Inert heterogeneities}
\label{sec:inert}

In the presence of $M$ inert spatial heterogeneities the transition probabilities between any two sites can be modified, leading to the Master equation \cite{sarvaharmangiuggioli2023}
\begin{align}
\mathcal{P}(\nvec, t + 1) &= \sum_{\mvec{m}}\mmat{A}_{\nvec, \mvec{m}}\, \mathcal{P}(\mvec{m},  t) \nonumber \\
&\qquad 
+ \sum_{k = 1}^{M}
\lb \delta_{\nvec, \vs_k}- \delta_{\nvec, \vps_k} \rb
\ls   \rjq[k] \mathcal{P}(\vs_k, t) \rd 
\ld- \ljq[k]\mathcal{P}(\vps_k, t) \rs,
\label{eq:rdprop_setup_full_master_main}
\end{align}
\begin{figure}[ht]
  \centering
\includegraphics[width=\textwidth]{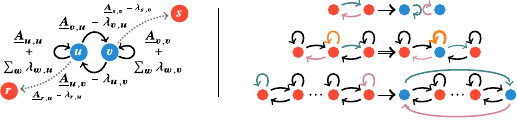}
\caption{Schematic representation of the transition probabilities after the introduction of spatial heterogeneities or disorder between a pair of sites subject to the constraints $\rjq[][\ws] \leq \matjump{\ws}{\vs}$ for all neighbouring sites $\ws$ with a (modified) heterogeneous connection in the direction
$\vs$ to $\ws$, and $0 \leq \matjump{\vs}{\vs} + \sum_{\ws} \rjq[][\ws]$ to ensure that the heterogeneous transition probabilities are positive. The left panel shows that the probability of hopping from site $\vs$ to $\vps$ is given by $\matjump{\vps}{\vs}$. When $\rjq$ is positive, the probability of jumping from $\vs$ to $\vps$ decreases, while the probability of staying put increases. When $\rjq$ is negative, the opposite
effect occurs with a decrease in the probability of staying, while
increasing the jump probability from $\vs$ to $\vps$. The parameter $\ljq$ affects the transition probability from $\vps$ to $\vs$ and the probability of remaining at $\vps$ in an equivalent
    manner. On the right panel we show three specific modifications: the introduction of a reflecting barrier (top), of a sticky site (middle), and of a long-range connection (bottom).}
  \label{fig:defect_schematic}
\end{figure}
where  $\mmat{A}_{\boldsymbol{n}, \boldsymbol{m}}$ is a compact notation for the tensor representing the transition probabilities between any pair of sites, e.g. Eq. (\ref{eq:bias_nd_master}) if the lattice is unbounded, or Eq. (\ref{eq:bias_nd_master}) together with the appropriate boundary constraints if the domain is finite. The role of the right elements inside the summation can be evinced by looking at Fig. \ref{fig:defect_schematic} where we illustrate how the transition probabilities between pairs of lattice sites are being modified whenever the $\lambda$ parameters are non-zero.

By generalising the so-called Montroll's defect technique for reactive sites to the case of inert heterogeneities \cite{sarvaharmangiuggioli2023}, we have obtained the exact propagator generating function of Eq. (\ref{eq:rdprop_setup_full_master_main}) as 
\begin{equation}
\widetilde{\mathcal{P}}_{\novec}(\nvec,z)=\propzz{\nvec}{\novec} - 1 + \frac{\mdet{\hcnmat[0]}}{\mdet{\hmat}},
  \label{eq:rdprop_sol_mt}
\end{equation}
where $\propzz{\nvec}{\novec}$ is the propagator generating function of the defect free problem, while $\mdet{\hmat}$ and
$\mdet{\hcnmat[0]}$ are determinants with matrices of size $M\times M$ built from the defect-free propagator dynamics and the heterogeneity properties as follows:
\begin{align}
  \label{eq:hmat_mt}
  &\hmat_{\matidx{i}{j}} = \rjq[i]  \propzzdif{\vs_i}{\vs_j}{\vps_j}
 - \ljq[i] \propzzdif{\vps_i}{\vs_j}{\vps_j} -  z^{-1}\delta_{i, j}, \\
  \label{eq:hcn0mat_mt}
  &\hcnmat[0]_{\matidx{i}{j}} = \hmat_{\matidx{i}{j}} - \propzzdif{\nvec}{\vs_j}{\vps_j}  
  \ls \rjq[i]\propzz{\vs_i}{\novec} - \ljq[i]
    \propzz{\vps_i}{\novec} \rs,
\end{align}
where $\propzzdif{\nvec}{\vs}{\vps}=\propzzdifful{\nvec}{\vs}{\vps}$ represents the difference between defect-free propagators evaluated at pairs of sites.
\begin{figure}[ht]
  \centering
  \includegraphics[width=\textwidth]{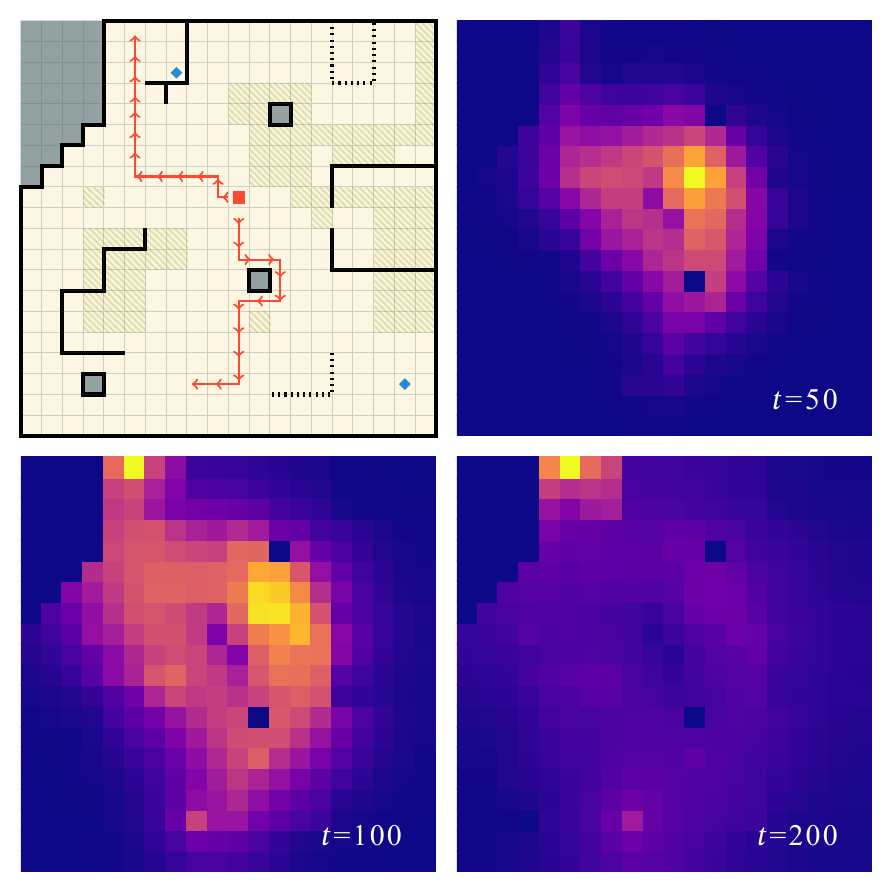}
  \caption{Spatial dependence of the propagator $\mathcal{P}_{\novec}(\nvec,t)$ at different times for a diffusive LRW starting from $\boldsymbol{n}_0=(11,12)$ in a square disordered lattice of size $N=20$ indicated in the left panel. The diffusivity along each axis is set to $q_i = 0.2$. The inert spatial heterogeneities are as follows: arrows indicate a local bias,  solid and dotted lines are, respectively, fully reflecting or permeable barriers, while shaded areas represent sticky sites. The permeable barriers are constructed such that 80\% of the time they reflect the walkers allowing the walker to pass through the remaining 20\% of the time, the sticky sites results in a reduction of movement by a factor of $0.2$, finally, the local biases are constructed to reduce the sojourning probability by 20\% whilst increasing the probability of moving in the indicated biased direction by the equivalent probability.  
  The blue marked diamond at coordinates (8,18) and (19,3) represent target sites used later in Sec. \ref{sec:fpp}.}
  \label{fig:disorder}
\end{figure}

We consider a disordered lattice represented in the left panel of Fig. \ref{fig:disorder}, and through a straightforward numerical $z$-inversion \cite{abatewhitt1992,abateetal2000} of Eq. (\ref{eq:rdprop_sol_mt}), we plot  $\mathcal{P}_{\novec}(\nvec,t)$ at times $t=50$, 100 and 200 in the right panels of Fig. \ref{fig:disorder}. It is evident that the local biases and the sticky sites force the occupation probability to deviate quite considerably from  a uniform spreading around the starting site even at relatively early times.

In some instances, when the number of spatial heterogeneities is large, we have devised a technique that avoids to calculate the determinants in Eq. (\ref{eq:rdprop_sol_mt}). This aspect is particularly convenient to study LRW dynamics in a space with multiple domains. For want of space we refer the reader to the publication \cite{dasgiuggioli2023} for further details. That mathematical technique has also allowed us to clarify the microscopic limiting procedure to obtain the so-called leather or Tanner boundary condition \cite{tanner1978} in continuous space, something relevant to study transport processes in porous materials.

\subsection{Resetting dynamics}
\label{sec:reset}

When a LRW is subject to resetting to a given site $\bm{n}_c$, its dynamics is as follows: at each time step it either resets to $\bm{n}_c$ with a given probability $r$ ($0\leq r\leq 1$), or performs its underlying Markov walk with the complementary probability $1-r$. The Master equation of the occupation probability of such a process, $_rP_{\bm{n}_0}(\bm{n},t)$ (the left subscript $r$ expresses the fact that the LRW is subject to resetting), can be written \cite{dasgiuggioli2022} in terms of the LRW dynamics in the absence of resetting. The renewal form of the Master equation is given by
\begin{align}
_rP_{\bm{n}_0}(\bm{n},t) =r\sum_{t'=0}^{t-1} (1-r)^{t'} P_{\bm{n}_c}(\bm{n},t')+(1-r)^t P_{\bm{n}_0}(\bm{n},t) \,.
\label{eq:Qreset-renewal}
\end{align}  
The structure of Eq. (\ref{eq:Qreset-renewal}) is advantageous as it connects directly the resetting propagator $_rP_{\bm{n}_0}(\bm{n},t)$ to the reset-free propagator $P_{\bm{n}_0}(\bm{n},t)$, thus allowing to express various quantities of the resetting dynamics in terms of the reset-free propagator. 

In $z$-domain, one may write Eq.~\eqref{eq:Qreset-renewal} as
\begin{align}
_r\widetilde{P}_{\bm{n}_0}(\bm{n},z) = \frac{r z}{1-z} ~ \widetilde{P}_{\bm{n}_c}(\bm{n},(1-r)z) +  \widetilde{P}_{\bm{n}_0}(\bm{n},(1-r)z) \,.\label{eq:Qreset_renewal_inz}
\end{align}
The form of Eq. (\ref{eq:Qreset_renewal_inz}) is particularly useful to determine the steady-state probability in the presence of resetting. In the limit $t \to \infty$, using the final value theorem for the $z$-transform, we obtain the steady-state probability $_r{P}^{\mathrm{ss}}(\bm{n})$ as
\begin{align}
_r{P}^{\mathrm{ss}}(\bm{n}) \equiv  \lim_{z \to 1}[(1-z)\, _r\widetilde{P}_{\bm{n}_0}(\bm{n},z) ] =  r \,\widetilde{P}_{\bm{n}_c}(\bm{n},(1-r)) \,, \label{eq:Qreset_SS}
\end{align}
which indicates, as expected, that the steady-state probability depends on the resetting site $\bm{n}_c$ and is independent of the initial site $\bm{n}_0$. For an unbiased 1$d$ LRW with diffusivity $q$ in unbounded domain subject to resetting at site $n_c$, the steady-state probability obtained using  Eq.~\eqref{eq:Qreset_SS} is given by 
\begin{align}
    _rP^{\mathrm{ss}}(n)  &=\frac{\sqrt{\chi_r-1}}{\sqrt{\chi_r + 1}} \exp{-|n-n_c| \ln(\chi_r+ \sqrt{\chi_r^2 - 1}) }, \label{eq:Qreset_SS_1D_unbiased}
\end{align}
where $\chi_r \equiv 1 + r / (q(1-r)) \geq 1$ is a renormalized parameter dependent only on the LRW diffusivity and the resetting probability.

We emphasise that Eqs.~\eqref{eq:Qreset-renewal},~\eqref{eq:Qreset_renewal_inz}, and~\eqref{eq:Qreset_SS} hold for any reset-free propagator and its generating function, and can thus be employed in arbitrary dimensions with arbitrary boundary conditions, e.g. for any choice of $\gamma$ in Eq. (\ref{bias:eq:full_nd_prop}), or with any spatial heterogeneity as dictated by Eq. (\ref{eq:rdprop_sol_mt}), as well as with arbitrary geometries, an example of which is presented in the next subsection.

\subsection{Hexagonal and honeycomb domains}\label{sec:hex}
It is well-known that it is not only the dimensionality of the lattice but its coordination number, that is the number of nearest neighbours of each site, which affects the properties of the walk \cite{zumofen1982energy}. This is exemplified in finite lattices where the combination of the shape or outer boundaries of the spatial domain and the coordination number have critical implications on the LRW dynamics. An important non-hypercubic lattice is the six neighbour hexagonal lattice employed in models ranging from the movement of mobile phone users \cite{akyildiz2000new} through to the formation of territories in scent-marking organisms \cite{giuggiolietal2011,giuggiolietal2013a,giuggiolikenkre2014,heiblumroblesgiuggioli2018,sarvaharmanetal2019}. In this section we introduce the periodic propagators for bounded hexagonal lattices and describe the wealth of quantities that can be obtained from them.
\begin{figure}
    \centering
\floatbox[{\capbeside\thisfloatsetup{capbesideposition={top, right},capbesidewidth=6cm}}]{figure}[\FBwidth]
{\caption{Schematic representation of the hexagonal lattice and the LRW movement steps. The $n_i$ unit vectors of the Her's coordinate system are shown next to the sites, while permissible movement directions between lattice sites are shown with the arrows emerging from the centre. The three outer numbers indicate the increase, decrease or no change of the $n_i$ coordinate along the specified direction. The magnitude of the transition probability along each direction, $(q/6)$, is modified when either of the three biases is non-zero, namely by the multiplicative factor $1\pm g_{i}$. For clarity, we omit arrows depicting the option of staying on lattice sites, which is given by $1-q$, and we also do not display the modifications that occur when periodic or reflecting boundaries are present.}}
{\includegraphics[width = 0.45\textwidth]{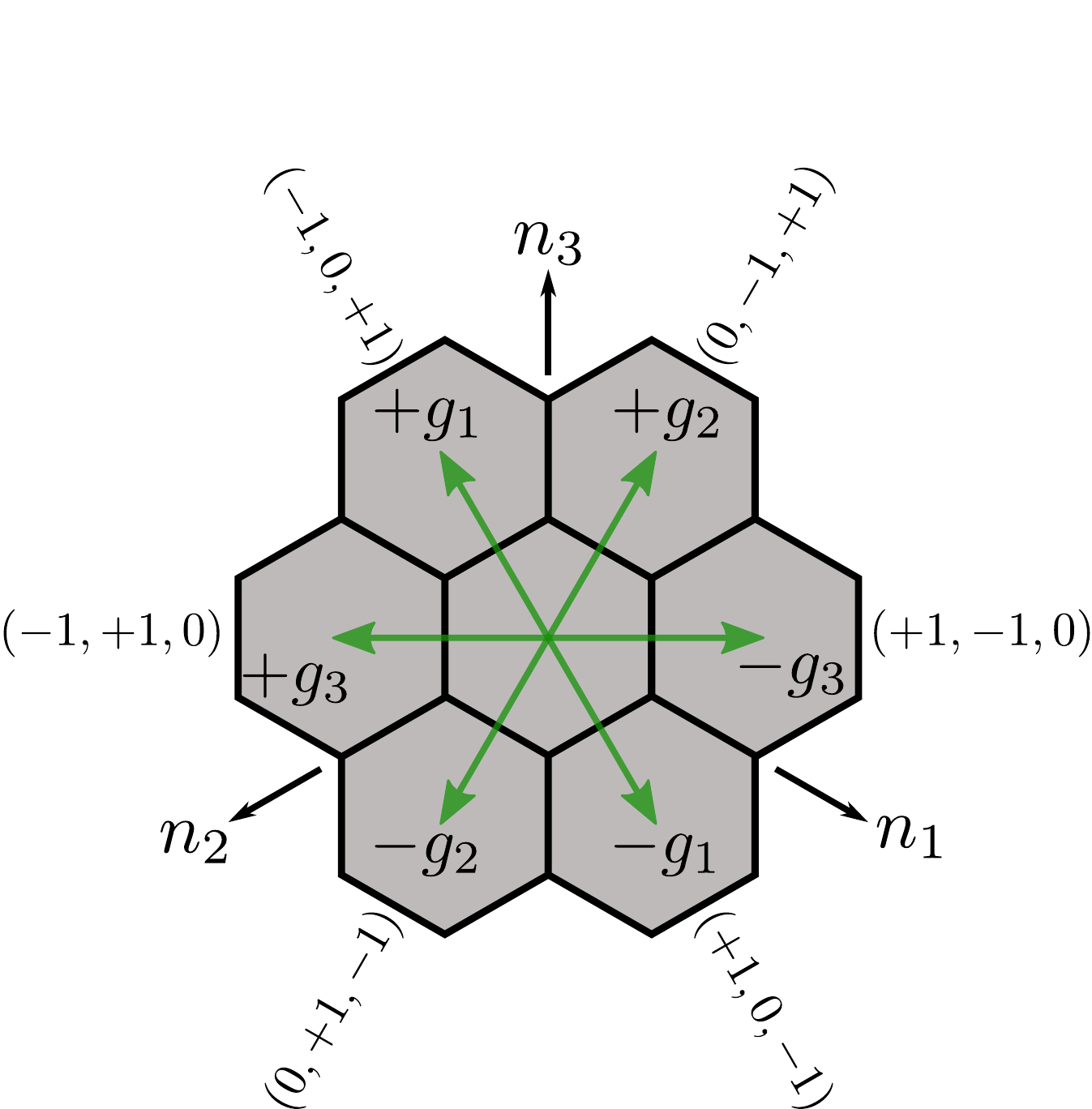}
\label{fig:hex_coords}}
\end{figure}

When considering hexagonal geometries, the non-orthogonality of the lattice sites makes it convenient to use the so-called Her's three-axis coordinate system \cite{her1994resampling, her1995geometric}, where each coordinate is given by three linearly dependent integer coordinates $(n_1, n_2, n_3)$ such that $n_1 + n_2 + n_3 = 0$ as displayed in Fig. \ref{fig:hex_coords}. 

Since the coordinates are linearly dependent, we require only two of them to uniquely describe a coordinate. In unbounded space, the Master equation governing the evolution of the occupation probability, $Q^{\mathcal{H}}(n_1, n_2, t)$ (the superscript $\mathcal{H}$ stands for hexagonal), is
\begin{equation}
\begin{aligned}
    Q^{\mathcal{H}}(n_1, n_2, &t+1) = \frac{q}{6}\Big[(1+g_1)Q^{\mathcal{H}}(n_1+1, n_2, t) +(1-g_1)Q^{\mathcal{H}}(n_1-1, n_2, t) \\ & + (1+g_2)Q^{\mathcal{H}}(n_1, n_2+1, t) +(1-g_2)Q^{\mathcal{H}}(n_1, n_2-1, t) \\ & + (1+g_3)Q^{\mathcal{H}}(n_1+1, n_2-1, t) +(1-g_3)Q^{\mathcal{H}}(n_1-1, n_2+1, t) \Big]\\ &+ (1-q)Q^{\mathcal{H}}(n_1, n_2, t),
\end{aligned}
    \label{eq: hex_master_equation}
\end{equation}
where $-1\leq g_{i} \leq 1$, $i \in \{1, 2, 3\}$ governs the independent bias along the three movement directions as shown in Fig. \ref{fig:hex_coords}. Following standard procedures \cite{hughesbook1995}, we obtain the unbounded propagator
\begin{equation}
\label{eq:unbounded_prop}   \widetilde{Q}^{\mathcal{H}}_{\bm{n}_{0}}(n_1, n_2, z) = \frac{1}{(2\pi)^2}\int_{-\pi}^{\pi} \int_{-\pi}^{\pi} \frac{\mathrm{e}^{i[(\bm{n} - \bm{n}_0)\cdot \bm{k}]}}{1-z \mu(k_1, k_2)}\mathrm{d}k_1\mathrm{d}k_2,
\end{equation}
where  $\bm{k} = (k_1, k_2)^{\intercal}$ (the symbol $^\intercal$ indicating the transpose operation), $\bm{n}-\bm{n}_0 = (n_{_1} - n_{0_1}, n_{_2} - n_{0_2})$ and
\begin{align}
\mu(k_1, k_2)
= 1-q &+\frac{q}{3}\Big[\cos(k_1) + \cos(k_2) + \cos(k_1-k_2) \Big.\nonumber \\
&\Big. - i g_1\sin(k_1)- i g_2\sin(k_2) - i g_3 \sin(k_1 - k_2)\Big],
\label{eq:mu}
\end{align}
is the so-called structure or characteristic function \cite{montrollweiss1965, klafter2011first}, or discrete Fourier transform of the step probabilities. The quantity $\widetilde{Q}^{\mathcal{H}}_{\bm{n}_{0}}(n_1, n_2, z)$ reduces to known results \cite{montroll1969} when $q = 1$ and $g_i = 0$, $i \in \{1, 2, 3\}$. We note here that in the other limit viz. $g_i = \pm 1$, $i \in \{1, 2, 3\}$, the walker is confined to three choices of movement direction. One can obtain a fully ballistic walker by solving a slightly modified Master equation whereby individual $q_i$ values are assigned to each unit vector such that $(q_1 + q_2 + q_3)/3 = q$. For example, in this case by setting $g_1 = 1$, $q_2= q_3 = 0$ one obtains ballistic motion along the $(-1, 0, +1)$ direction.

In order to find the exact dynamics in finite hexagonal domains, we make use of the so-called method of images, a technique exploited extensively for LRW in hypercubic lattices. For the technique to be applied to hexagonal domains, a generalisation to non-orthogonal coordinate systems is necessary. In doing so for periodic hexagonal domains, as a result of the zig-zag nature of the boundaries there are two unique image sets, which we refer to as the left and right shift (see Figs. 2 and 8 in ref. \cite{marris2023exact} for pictorial representations). As the two image sets create two unique periodic propagators with differing transport statistics, we denote them below with the superscripts $\ell$ and $\mathfrak{r}$, respectively.

In applying the generalised method of images we take a finite hexagon with a circumradius of size $R$, containing $\Omega = 3R^2 + 3R + 1$ lattice sites, and we construct the periodic propagator via \cite{marris2023exact} $\widetilde{P}^{\mathcal{H}^{(p)}}_{\bm{n}_{0}}(n_1, n_2, z) = \sum_{m_1 = -\infty}^{\infty}\sum_{m_2 = -\infty}^{\infty}\widetilde{Q}^{\mathcal{H}}_{\bm{n}_{0}}(n_1 + \hat{n}^{[i]}_1, n_2 + \hat{n}^{[i]}_2, z) $ with the $[i]$ shift using
\begin{equation}
\label{eq:LRshifts}
    \begin{bmatrix}
        \widehat{n}^{[\ell]}_1 \\[1ex] \widehat{n}^{[\ell]}_2 
    \end{bmatrix} = \begin{bmatrix}
        (2R+1)m_1 -R m_2 \\  -(R+1)m_1 + (2R+1)m_2
    \end{bmatrix};  \; \: \;
    \begin{bmatrix}
        \widehat{n}^{[\mathfrak{r}]}_1 \\[1ex] \widehat{n}^{[\mathfrak{r}]}_2 
    \end{bmatrix} = \begin{bmatrix}
        -Rm_1 +(2R+1)m_2 \\ -(R+1)m_1 - Rm_2
    \end{bmatrix}.
\end{equation}
Using (\ref{eq:LRshifts}), we find the periodic propagator in closed form as
\begin{equation}
\begin{aligned}
P_{\bm{n}_0}^{\mathcal{H}^{(p)_{[i]}}}(n_1, n_2, t) = \frac{1}{\Omega} &+ \frac{1}{\Omega}\sum_{r=0}^{R-1}\sum_{s=0}^{3r+2}\bigg\{\mathrm{e}^{i\left[(\bm{n} - \bm{n}_0)\cdot \bm{k}^{[i]}\right]}\left[\mu\left(k_1^{[i]}(r,s), k_2^{[i]}(r,s)\right)\right]^t  \\ &
+ \mathrm{e}^{-i\left[(\bm{n} - \bm{n}_0)\cdot \bm{k}^{[i]}\right]}\left[\mu\left(-k_1^{[i]}(r,s), -k_2^{[i]}(r,s)\right)\right]^t \bigg\},
\end{aligned}
\label{eq:hexagonal_periodic}
\end{equation}
where $k^{[\mathfrak{r}]}_1(r,s) = k^{[\ell]}_2(r,s) = 2\pi \Omega^{-1} \left[R(s+1) + s - r\right]$  and $k^{[\mathfrak{r}]}_2(r,s) = k^{[\ell]}_1(r,s) $  $= 2\pi \Omega^{-1} [R(2-s+3r) + r + 1]$.
Equation (\ref{eq:hexagonal_periodic}) reduces to Eq. (12) of ref. \cite{marris2023exact} in the absence of any bias. Note that using Eq. (\ref{eq:hexagonal_periodic}) the LRW propagator in a finite hexagons with reflecting boundaries can also be found by using the general framework for spatial heterogeneities presented in Sec. \ref{sec:inert}. To do that one needs to insert $6R+3$ pairs of defective sites (3 pairs when the site is a corner site, and 2 pairs otherwise) to ensure that a walker cannot jump to periodically-linked hexagonal boundary sites on the other side of the lattice. Such an approach was used for the unbiased case and can be found in ref. \cite{marris2023exact}.

The creation of the hexagonal periodic image set in Eq. \ref{eq:LRshifts} has also opened up the ability to represent, entirely analytically, random motion in six-sided hexagonal geometry. We note in fact that Eq. (\ref{eq:hexagonal_periodic}) is valid with any normalised structure-function, and the inclusion of internal states \cite{weiss1983random, roerdink1985asymptotic} into the random walk allows the analytic representation of the three-neighbour, non-Bravais honeycomb lattice \cite{marris2023exact}, which finds use in transport processes over carbon structures such as graphite \cite{bercu2021asymptotic} and single-walled carbon nanotubes \cite{cotfas2000random, batchelorhenry2002}. We omit the details here, but we point the interested reader once again to ref. \cite{marris2023exact}.

\section{First passage processes to multiple targets}
\label{sec:fpp}

The LRW renewal equation, first introduced by Erd\"{o}s, Feller and Pollard \cite{erdosetal1949}, and used extensively since (see e.g. \cite{redner2001,benichouvoituriez2014,metzleretal2014b}), has a fundamental role in first-passage processes since it allows to link the generating functions of the occupation probability with the first-passage probability to reach the target site $\boldsymbol{n}$ from the initial site $\boldsymbol{n}_0$:
\begin{equation}
\widetilde{F}_{\boldsymbol{n}_0\rightarrow \boldsymbol{n}}(z)=\frac{\widetilde{P}_{\boldsymbol{n}_0}(\boldsymbol{n},z)}{\widetilde{P}_{\boldsymbol{n}}(\boldsymbol{n},z)}.
\label{eq:renewal}
\end{equation}

Generalisations of relation (\ref{eq:renewal}) to multiple targets have appeared, e.g. \cite{larraldeweiss1995,chevalieretal2011,giuggioli2020}, but the proposed relations have been constructed in terms of the solution of an algebraic set of coupled probability equations. In \cite{rubinweiss1982} Rubin and Weiss developed a LRW formalism to compute the number of visits to an arbitrary set of lattice sites in unbounded or periodic domains. The generalisation of that method to arbitrary geometry, devised in ref. \cite{giuggiolisarvaharman2022}, has led recently to the sought-after general theory of first-passage processes to a set of multiple (partially) absorbing targets.

With a set of targets at coordinates $\boldsymbol{m}_i$ $(i=1,...,M)$ and  absorbing probability $\boldsymbol{\rho}=\big(\rho_{\boldsymbol{m}_1},\rho_{\boldsymbol{m}_2},...,\rho_{\boldsymbol{m}_M}\big)$ with ($0<\rho_i\leq 1$),
the multi-target theory allows to determine first-passage probabilities to any of the absorbing sites as well as the so-called splitting probabilities to reach only one, that is the probability to reach target $j$ and none of the remaining ones, $\widetilde{F}_{\boldsymbol{n}_0\rightarrow\left(\boldsymbol{m}_j|\boldsymbol{m}_1;...;\boldsymbol{m}_{j-1};\boldsymbol{m}_{j+1};...;\boldsymbol{m}_M\right)}(\boldsymbol{\rho},z)$. That quantity is given in terms of individual first-passage and first-return probability as follows 
\begin{equation}
\widetilde{F}_{\boldsymbol{n}_0\rightarrow\left(\boldsymbol{m}_j|\boldsymbol{m}_1;...;\boldsymbol{m}_{j-1};\boldsymbol{m}_{j+1};...;\boldsymbol{m}_M\right)}(\boldsymbol{\rho},z)=\frac{|\mathbb{F}^{(j)}(\boldsymbol{\rho},z)|}{|\mathbb{F}(\boldsymbol{\rho},z)|},
\label{eq:split}
\end{equation}
where the $M\times M$ matrix $\mathbb{F}(\boldsymbol{\rho},z)$ has 
diagonal components equal to $\mathbb{F}_{k,k}(\boldsymbol{\rho},z)=1+\frac{1-\rho_{\boldsymbol{m}_k}}{\rho_{\boldsymbol{m}_k}}\left[1-\widetilde{\mathcal{R}}_{\boldsymbol{m}_k}(z)\right]$, with $\widetilde{\mathcal{R}}_{\boldsymbol{m}_k}(z)$ the generating function of the (first-)return probability to site $\boldsymbol{m}_k$, and the off-diagonal terms $\mathbb{F}_{j,k}(\boldsymbol{\rho},z)=\widetilde{F}_{\boldsymbol{m}_k\rightarrow \boldsymbol{m}_j}(z)$. The matrix $\mathbb{F}^{(j)}(\boldsymbol{\rho},z)$ is the same as $\mathbb{F}(\boldsymbol{\rho},z)$, but with the $j$-th column replaced by $\mathcal{G}(z)=\big(\widetilde{F}_{\boldsymbol{n}_0\rightarrow\boldsymbol{m}_1}(z),\widetilde{F}_{\boldsymbol{n}_0\rightarrow\boldsymbol{m}_2}(z), \ldots ,\widetilde{F}_{\boldsymbol{n}_0\rightarrow\boldsymbol{m}_M}(z)\big)^{\intercal}$.

If one were interested in the first-passage over a given subset of the $M$ targets, given the mutually exclusive nature of the first-passage and first-return trajectories, it suffices to sum Eq. (\ref{eq:split}) over the appropriate $j$ elements. When the interest instead lies in knowing the first-passage probability to any of the $M$ targets it is trivial to show that the multi-linearity of the determinant allows to write the first-passage probability to any of the targets as
\begin{equation}
\widetilde{F}_{\boldsymbol{n}_0\rightarrow\left(\boldsymbol{m}_1;...;\boldsymbol{m}_M\right)}(\boldsymbol{\rho},z)= 1- \frac{|\overline{\mathbb{F}}(\boldsymbol{\rho},z)|}{|\mathbb{F}(\boldsymbol{\rho},z)|},
\label{eq:split_all}
\end{equation}
where $\overline{\mathbb{F}}_{i,j}(\boldsymbol{\rho},z)=\mathbb{F}_{i, j}(\boldsymbol{\rho},z)-\mathcal{G}_{i}(z)$.

Using Eq. (\ref{eq:split_all}) the mean first-passage, or mean first-absorption time when $\rho<1$, to any of the $M$ targets can be conveniently obtained by standard procedure, namely $T_{\boldsymbol{n}_0\rightarrow\left(\boldsymbol{m}_1;...;\boldsymbol{m}_M\right)}=\left.(\mathrm{d}/\mathrm{d}z)\,\widetilde{F}_{\boldsymbol{n}_0\rightarrow\left(\boldsymbol{m}_1;...;\boldsymbol{m}_M\right)}(\boldsymbol{\rho},z)\right|_{z=1}$. The resulting expression is given by
\begin{equation}
T_{\boldsymbol{n}_0\rightarrow\left(\boldsymbol{m}_1;...;\boldsymbol{m}_M\right)}=\frac{|\mathbb{T}_0|}{|\mathbb{T}_1|-|\mathbb{T}|}.
\label{eq:MTT}
\end{equation}
The $M\times M$ elements of the matrix $\mathbb{T}$ are constructed by calculating individual mean first-passage and first-return quantities, namely, $\mathbb{T}_{i,i}=-\frac{1-\rho_{\bm{m}_i}}{\rho_{\bm{m}_i}}R_{\bm{m}_i}$ for the diagonal terms, with  
$R_{\bm{m}_i}$ the mean return time to $\bm{m}_i$, and the off-diagonal terms $\mathbb{T}_{i,j}=T_{\bm{m}_j\rightarrow \bm{m}_i}$, with the mean first-passage time from $\bm{m}_j$ to $\bm{m}_i$. The other matrices, $\mathbb{T}_0$ and $\mathbb{T}_1$, are constructed as follows: $\mathbb{T}_{0_{i,j}}=\mathbb{T}_{i,j}-T_{\bm{n}_0\rightarrow \bm{m}_i}$ and $\mathbb{T}_{1_{i,j}}=\mathbb{T}_{i,j}-1$.

As a transmission event of a token of information, e.g. the passing of a pathogen or an internal degree of freedom, or its limiting case an encounter event when $\rho=1$, is exactly equivalent to a first-absorption event at any of the locations where two LRW may meet \cite{kenkre1980}, an important achievement of Eq. (\ref{eq:split_all}) is that it allows to compute the time dependence of the probability of encounter and transmission \cite{giuggiolisarvaharman2022}. The necessary ingredients consist of the joint time-dependent occupation probability in the $2d$-dimensional space  for the two independent walkers that move in $d$ dimensions, which are now available in hypercubic lattices for simple \cite{giuggioli2020} and biased walks \cite{sarvaharmangiuggioli2020} with resetting dynamics \cite{dasgiuggioli2022,dasetal2023} as well as in triangular and hexagonal lattices \cite{marris2023exact}. Using the knowledge of the joint propagators, the evaluation of mean first-passage times between lattice points and mean return times is also possible, leading to the evaluation of the mean transmission or mean encounter time via Eq. (\ref{eq:MTT}).

\section{Dynamics in the presence of radiation boundaries}
\label{sec:rad}

The so-called radiation boundary condition, introduced in LRW theory by Weiss \cite{weiss1994}, consists of a lattice site at which a walker may be reflected with probability $\eta$ and absorbed with probability $1-\eta$. If we consider for simplicity a 1D unbounded LRW, described by Eq. (\ref{eq:bias_nd_master}) in the absence of any bias, starting to the right of the origin and such that a radiation boundary is present at the origin, the boundary constraint imposes the following flux condition \cite{kaygiuggioli2023}:
\begin{equation}
\label{eq:flux_condition}
    J^+(0,t)=\eta J^-(0,t),
\end{equation}
where the $-$ and $+$ superscripts indicate the flux in and out of the origin, respectively, while the total flux is given by $J(0,t)=J^+(0,t)-J^-(0,t)$. If we take a nearest-neighbour 1$d$ LRW, with lattice points $n=0,1,...$, the fluxes in and out of the origin are, $J^-(0,t)=q P(1,t)/2$ and $J^+(0,t)=q P(0,t)/2$. These relations lead to the discrete version of the so-called radiation (also referred to as Robin or Fourier) boundary condition
\begin{equation}
\label{eq:discrete_radiation}
J(0,t)=\frac{q(\eta-1)}{2\eta} P(0,t),
\end{equation}
which is a generalisation to the case of lazy LRWs, and are well-known in the literature (see e.g. \cite{grebenkovetal2003b}).
For $\eta=1$ we recover from Eq. (\ref{eq:discrete_radiation})  the reflecting boundary condition, whereas for $\eta=0$ we have the absorbing boundary condition.

To find the first-reaction (or first-absorption) probability when the LRW Master equation is supplemented by the boundary condition  (\ref{eq:discrete_radiation}), we need to formulate the dynamics  in terms of the number of times, $\mathcal{S}_t$, the LRW visits the origin \cite{it1965diffusion,freidlin1985functional,grebenkov2020,grebenkov2020b}. Using $\mathcal{N}_t$ as the stochastic variable that represents the location of the LRW, with $\mathcal{N}_0\neq0$, we define
$\mathcal{S}_t=\sum_{i=1}^t \delta_{\mathcal{N}_i,0}$ and write a Master equation describing the joint probability of $\mathcal{N}_t$ and $\mathcal{S}_t$, $\Lambda(n,s,t)=\mathbb{P}[\mathcal{N}_t=n,\mathcal{S}_t=s]$,  as 
\begin{align}
\label{eq:master_visits}
\Lambda(n,s,t+1) &=\left[\frac{q}{2}\Lambda(n-1,s,t)+\frac{q}{2}\Lambda(n+1,s,t)+(1-q)\Lambda(n,s,t)
    \right](1-\delta_{n,0}) \nonumber \\
    &\hspace{-1.5cm} +\left[\frac{q}{2}\Lambda(n-1,s-1,t)+\frac{q}{2}\Lambda(n+1,s-1,t)+(1-q)\Lambda(n,s-1,t)\right]\delta_{n,0}.
\end{align}
Taking the generating function of Eq. (\ref{eq:master_visits}) with respect to $s$, i.e. $\sum_{s=0}^{\infty} \eta^s\Lambda(n,s,t)=
P^{(\eta)}(n,t)$, we find
\begin{align}
\label{eq:discrete_Feynman_Kac}
    P^{(\eta)}(n,t+1)&=\left[\frac{q}{2}P^{(\eta)}(n-1,t)+\frac{q}{2}P^{(\eta)}(n+1,t)+(1-q)P^{(\eta)}(n,t)\right] (1-\delta_{n,0})  \nonumber \\
    &\hspace{-0.5cm} +\eta\left[\frac{q}{2}P^{(\eta)}(n-1,t)+\frac{q}{2}P^{(\eta)}(n+1,t)+(1-q)P^{(\eta)}(n,t)\right]\delta_{n,0} \nonumber \\
    &\hspace{-0.5cm} = {\eta}^{\delta_{n,0}} \left[ \frac{q}{2}P^{(\eta)}(n-1,t)+\frac{q}{2}P^{(\eta)}(n+1,t) +(1-q)P^{(\eta)}(n,t)\right],
\end{align}
which is the discrete time Feynman-Kac equation \cite{kac1949,kac1951,csaki1993} for the functional that indicates the number of visit to the origin.  Through a simple rearrangement of the terms on the right hand side of the top equation in (\ref{eq:discrete_Feynman_Kac}), one realises that we have written the Master equation of a lazy symmetric unbounded LRW with a partially absorbing defect at $n=0$ with absorption probability $\rho=1-\eta$.

Since $\Lambda(n,s,t)=\left\langle \delta_{\mathcal{N}_t,n} \  \delta_{\mathcal{S}_t,s} \right\rangle$, we have that the $\eta$-transform is $P^{(\eta)}(n,t)=\sum_{s=0}^\infty \eta^s \Lambda(n,s,t)=\left\langle \delta_{\mathcal{N}_t,n} \ {\eta}^{\mathcal{S}_t} \right\rangle,$
which leads to 
\begin{equation}
    P^{(\eta)}(n,t)=\sum_{s=0}^\infty {\eta}^s \Big(\mathbb{P}[\mathcal{N}_t=n,\mathcal{S}_t<s+1]-\mathbb{P}[\mathcal{N}_t=n,\mathcal{S}_t<s]\Big).
\end{equation}
As $\mathbb{P}[\mathcal{N}_t=n,\mathcal{S}_t<0]=0$,
we may write
\begin{equation}
   P^{(\eta)}(n,t)=\sum_{s=1}^\infty \mathbb{P}[\mathcal{N}_t=n,\mathcal{S}_t< s](\eta^{s-1}-{\eta}^s)=\sum_{s=0}^\infty \mathbb{P}[\mathcal{N}_t=n,\mathcal{S}_t\leq s](1-\eta)\eta^s.
\end{equation}
Identifying $(1-{\eta}){\eta}^s$ as the probability distribution of a geometrically distributed random variable, $\hat{s}$, i.e. $\mathbb{P}[\hat{s}=s]=\langle \delta_{\hat{s},s} \rangle = (1-\eta)\eta^s$, we obtain
\begin{equation}\label{eq: rad_sol}
    P^{(\eta)}(n,t)= \left\langle \sum_{s=0}^\infty \mathbb{P}[\mathcal{N}_t=n,\mathcal{S}_t\leq s] \delta_{\hat{s},s}\right \rangle =\mathbb{P}[\mathcal{N}_t=n,\mathcal{S}_t\leq \hat{s}].
\end{equation}
Thus, $P^{(\eta)}(n,t)$, the solution of the Feynman-Kac equation (\ref{eq:discrete_Feynman_Kac}), represents the LRW occupation probability at $n$, whilst having not visited the origin more times than a geometrically distributed random variable, $\hat{s}$, with mean $\eta/(1-\eta)$.

Now, we return to the radiation boundary. Consider a random walker that visits the origin $m$ times without being absorbed, but being reflected at each visit, then the probability of this is simply $\eta^m$, $\eta$ being the reflection probability. As the number of visits to the origin is a random variable, $\mathcal{S}_t$, we must average over all possible realisations of $\mathcal{S}_t$. The probability for the random walker to be located at the lattice point $n$ at time $t$, whilst having not been absorbed by the radiation boundary, is $\left\langle \delta_{\mathcal{N}_t,n} \ {\eta}^{\mathcal{S}_t} \right\rangle$, which is simply the solution to Eq. (\ref{eq:discrete_Feynman_Kac}), with a reflecting boundary condition at the origin. Thus we have established that the radiation boundary can be formulated in terms of a reflected random walk that is absorbed once the number of visits to the origin exceed a geometrically distributed random variable with mean $\eta/(1-\eta)$.

Following the arguments of Ref. \cite{grebenkov2020}, as $\mathcal{S}_t$ is a monotonically non-decreasing process, the event $\left\{\mathcal{S}_t>\hat{s}\right\}$ is equivalent to $\left\{t>\mathcal{T}\right\}$, where $\mathcal{T}$ is the time of reaction (absorption), therefore 
\begin{equation}\label{eq:react_time}
    \mathcal{T}=\min\left\{t>0: \mathcal{S}_t>\hat{s} \right\}.
\end{equation}
Equation (\ref{eq:react_time}) represents the spatio-temporal discrete form of the reaction time, which was found in the continuous paradigm in Ref. \cite{grebenkov2020}. From Eq. (\ref{eq:react_time}) one obtains
\begin{equation}\label{eq:prop_react}
P^{(\eta)}(n,t)=\mathbb{P}\left[\mathcal{N}_t=n,t<\mathcal{T} \right].
\end{equation}
One is then able to study the first-reaction time probability, $\mathcal{F}^{(\eta)}_{n_0\rightarrow 0}(t)=\mathbb{P}[\mathcal{T}=t|\mathcal{N}_0=n_0]$, for a random walker in the presence of a radiation boundary. Due to the connection between the radiation boundary and discrete Feynman-Kac equation explained above, we solve Eq. (\ref{eq:discrete_Feynman_Kac}), with a reflecting boundary at the origin and the localized initial condition at $n_0$, using the defect technique. The resulting probability is the propagator of Eq. (\ref{eq:bias_nd_master}) for a 1$d$ unbiased walker that satisfy the boundary condition (\ref{eq:discrete_radiation}).

Then marginalisation over $n$ leads to the survival probability, $\mathbb{P}[t<\mathcal{T}]$, which then gives the first-reaction probability as
\begin{equation}
\widetilde{\mathcal{F}}^{(\eta)}_{n_0\rightarrow 0}(z)=\frac{\widetilde{P}_{n_0}(0,z)+\widetilde{P}_{-n_0}(0,z)}{\frac{\eta}{1-{\eta}}+2\widetilde{P}_{0}(0,z)},
\label{eq:freact}
\end{equation}
where $\widetilde{P}_{n_0}(n,z)$ is the generating function of the lazy 1$d$ symmetric unbounded propagator \cite{giuggioli2020}, i.e.
\begin{equation}
    \widetilde{P}_{n_0}(n,z)=\frac{\left(\frac{1}{\beta(z)}+\frac{1}{\beta(z)}\sqrt{1-\beta^2(z)}\right)^{-|n-n_0|}}{\left[1-z(1-q)\right]\sqrt{1-\beta^2(z)}},
\end{equation}
where $\beta(z)=zq/[1-z(1-q)]$ with $|\beta(z)|\leq 1$ (see Sec. \ref{sec:fpfeft} for the time dependence of $\mathcal{F}^{(\eta)}_{n_0\rightarrow 0}(t)$).

\section{Dynamics of first-passage, first-encounter and first-reaction processes}
\label{sec:fpfeft}

\begin{figure}[t]
\centering
\includegraphics{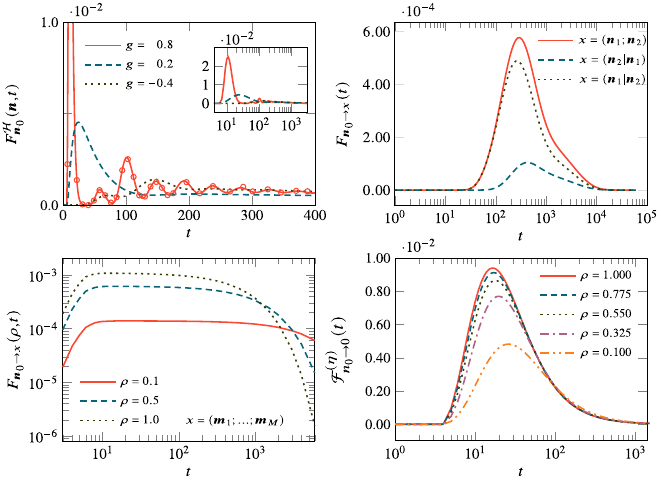}
\caption{Top left panel: First passage time from the origin to a single target at $(n_1, n_2, n_3) = (7, -2, -5)$ in a left shift periodic hexagon with $R = 14$  for different bias strength $g$ with $g_i=g$ and diffusivity $q=0.9$. The solid lines is obtained by numerically inverting Eq. (\ref{eq:renewal}) and making use of Eq. (\ref{eq:hexagonal_periodic}), while the open-circle markers are results from $10^6$ stochastic simulations. Inset shows the same information plotted up to $t=3000$ on a semilog scale. Top right panel: First-passage probabilities to either of two targets as well as their splitting probabilities for the 2D LRW moving in the disordered domain shown in the left panel of Fig. \ref{fig:disorder}. The LRW parameters are the same as those in that panel, and the targets, indicated in blue, are located, respectively, at $\boldsymbol{n}_1=(8,18)$ and $\boldsymbol{n}_2=(19,3)$. The splitting probability to reach $\boldsymbol{n}_1$ is the curve with the higher mode. The propagator in the disordered space,  shown in Eq. (\ref{eq:rdprop_sol_mt}), is used in Eqs. (\ref{eq:split}) and (\ref{eq:split_all}) to plot, respectively, the splitting probabilities or the first-passage probability to either target. Bottom left panel: First-transmission probability for two independent resetting 1D LRW without bias in a periodic lattice of size $N=21$ ($1\leq n\leq 21$). The joint propagator for the two resetting walkers used in Eq.~\eqref{eq:split_all} is obtained by taking the generating function of the product of 1D time-dependent resetting propagators through Eq. (\ref{eq:Qreset-renewal}). The first and the second walker starts from their corresponding resetting sites set at $n_{c_1}=8$ and $n_{c_2}=14$, respectively. The diffusivities and resetting probabilities of the walkers are taken to be $q_1=q_2 = 0.8$ and $r_1 = r_2 =0.4$, respectively. The walkers can interact with probability $\rho$ when they simultaneously occupy a site $n$ in the 1D domain, which corresponds to a partially absorbing site $\boldsymbol{m}_i=(i,i)$, where $i \in [1,M=21]$ in the 2D domain. Bottom right panel: First-reaction probability, $\mathcal{F}^{(\eta)}_{n_0\rightarrow 0}(t)$, for a radiation boundary at the origin, for varying reaction probabilities $\rho=1-\eta$, for $n_0=5$ and $q=0.5$. The case $\rho=1$ or $\eta=0$ corresponds to the fully absorbing case in the flux and boundary condition, respectively, Eqs. (\ref{eq:flux_condition}) and (\ref{eq:discrete_radiation}), transforming the first-reaction probability into a first-passage probability. The expression for $\widetilde{\mathcal{F}}^{(\eta)}_{n_0\rightarrow 0}(z)$ can be found in Eq. (\ref{eq:freact}).}
\label{fig:compositeFP}
\end{figure}

To show the general applicability of the multi-target formalism described in Sec. \ref{sec:fpp}, we use it to plot the time dependent probability of various target hitting processes, namely a first-passage to a single target in a hexagonal domain, a first-passage to either of two targets in a disordered lattice, and a first-transmission between two resetting walkers in a periodic domain. We also exploit the formalism introduced in Sec. \ref{sec:rad} to display the dynamics of first-reaction to a single target in a 1D unbounded domain. We present these four examples in the four panels of Fig. \ref{fig:compositeFP}.

The top left panel displays a very rich first-passage dynamics with multimodal shapes, a feature rarely observed with Markov LRW, but present in non-Markov \cite{verechtchaguina2006first} and quantum walks \cite{kulkarni2023first}. This top left panel represents the first passage of a biased periodic hexagonal LRW from the origin to a target located towards the bottom right of the domain for different bias values. As the level of bias towards the target increases we see that the mode of $F^{\mathcal{H}}_{\boldsymbol{n}_0\rightarrow \boldsymbol{n}}(t)$ increases in magnitude and moves towards shorter times. With high bias, exemplified by the case with $g_{i} = 0.8$, the initial peak is followed by a region of near zero probability, corresponding to the timescale at which the walkers travelling towards the target are likely to have passed it and the walkers who originally wandered in the opposite direction have yet to reach the target. In general we observe that higher biases lead to sustained oscillations at intermediate times in $F^{\mathcal{H}}_{\boldsymbol{n}_0\rightarrow \boldsymbol{n}}(t)$, with troughs corresponding to timescales in which heavily biased trajectories are away from the target. Oscillations are also sustained for longer times for larger domain sizes as the chance of missing the target increases with the circumradius $R$. Eventually, once the indirect trajectories dominate the dynamics, the oscillations dampen and $F^{\mathcal{H}}_{\boldsymbol{n}_0}(\boldsymbol{n},t)$ drops to zero, as shown in the inset of the top left panel.

In the top right panel, we plot the the first passage and
 time-dependent splitting probabilities
 to either of the targets with the environment and initial
 condition shown in the left-most panel of  \cref{fig:defect_schematic} where the two targets, are indicated using blue
diamonds. Due to the local biases, the walker starting at $\boldsymbol{n}_0=(11,12)$ is carried quickly to
 the target $\nvec_1 = (8, 18)$ resulting in an increase of the splitting
 probability $F_{\novec \to \lb \nvec_1 | \nvec_2 \rb}(t) $ at earlier times. The local biases instead do not facilitate the movement towards the target at $\boldsymbol{n}_2=(19,3)$, which is reached mainly diffusively from $\boldsymbol{n}_0$. The
 reflecting barriers also act to confine the walker temporarily
 nearby $\nvec_1$ further increasing the likelihood of absorption
 at $\nvec_1$, while much of the heterogeneities
 present do not assist in reaching $\nvec_2$, and the walker must explore a much larger proportion of the domain before reaching $\nvec_2$, which results in a mode at a
 later time for $F_{\novec \to (\nvec_2 | \nvec_1)}(t)$. Beyond the delay in the mode of the splitting probabilities, another interesting feature is that the trajectories that contribute to the first-passage probability to either of the targets is mostly made up of trajectories that reach $\nvec_1$, as one may evince by the small difference in the curves of $F_{\novec \to (\nvec_1 | \nvec_2)}(t)$ and $F_{\novec \to (\nvec_1;\nvec_2)}(t)$. More quantitatively, one has that the relative ratio between the
 integrated splitting probabilities, i.e. $\sum_{t = 0}^{\infty}
F_{\novec \to (\nvec_1 | \nvec_2)}(t) \approx 0.74$ and 
$\sum_{t = 0}^{\infty} F_{\novec \to (\nvec_2 | \nvec_1)}(t) \approx
 0.26$. Given that
 in the absence of any heterogeneities, the proportion of trajectories that reach either $\nvec_1$ or $\nvec_2$ from $\novec$ is 
 more evenly split, these findings suggest empirical approaches to favour the selection of certain targets versus others based on the type and strength of spatial disorder that one may introduce in a given spatial domain.
 

In the bottom panels we show the first-transmission probability between two resetting LRWs (left) and the first-reaction probability at a lattice site for an individual LRW (right). In the left panel we show the effects of the co-location or proximity transfer efficiency, $\rho$, on the first-transmission probability (at any of the sites). The transport limited regime, which corresponds to the case $\rho=1$, is simply the encounter probability. It displays the highest mode and at the earliest time. The reaction limited regime is particularly evident when $\rho=0.1$, for which the curve remains essentially constant after the mode, with the flat profile lasting longer the smaller is $\rho$. Given that the system is finite eventually a fast decay to zero characterises all of the curves for any $\rho>0$. In the right panel, we present the first-reaction probability, $\mathcal{F}_{n_0\rightarrow 0}^{(\eta)}(t)$, for different reactivity, $\rho$, at the origin. One can clearly see that, as the reactivity gets reduced, the mode shifts to longer times as more and more visits to the origin are necessary before the walker reacts. The widening of the shape of the probability and the appearance of longer tails is also a result of the need for multiple return to the origin before a reaction takes place.

\section{Disorder indifference phenomena}
\label{sec:dis_indif}

When a LRW moves on a disordered lattice, the type and location of the spatial heterogeneities heavily affect the spatio-temporal dependence of the walker's occupation probability. But in some cases, certain quantities, surprisingly, are not affected at all by the disorder. We have uncovered two such examples and we report them below.

The first example is the mean first-passage time from site $n_0$ to site $n$ in a semi-bounded 1D domain in presence of a symmetric permeable barrier at lattice site $u$ in between $n$ and $n_0$. To be more precise consider a left reflecting barrier between site $n=0$ and site $n=1$, then take the LRW initial position at $n_0\geq 1$, and to the right of $n_0$ we place the barrier at $u$ and further to the right we have the site $n$, that is $1\leq n_0<u<n$. In this arrangement, while the time-dependence of the first-passage probability from $n_0$ to $n$ does depend on the location $u$ and the permeability of the barrier, which is proportional to the $\lambda$ parameter in Sec. \ref{sec:inert}, the mean first-passage time is independent of both. This surprising effect, reported in ref. \cite{sarvaharmangiuggioli2023}, which was shown to appear even in the presence of a global bias, is lost when the barrier permeability is made asymmetric. An analogous effect was also shown for Brownian walks \cite{kaygiuggioli2022} in 1D and in radially symmetric geometries \cite{godecmetzler2015}.

The second disorder indifference phenomenon we have observed manifests when one is interested in counting the number of visits or more generally the number of detectable visits. By detectable it is meant that there is a parameter governing the probability that the visit is accounted for. To define the problem more precisely, we consider the number of detectable visits that occurs at the set of sites $S_v$ before being
absorbed at any of the set of sites $S_a$.  Let the set $S_a = {\rss[1], \cdots, \rss[M_a]}$ be
the destination points where the process terminates, i.e. the targets, and let
$S_v = {\rss[M_a +1 ], \cdots, \rss[M_a + M_v]}$ be the set of sites or points of interest where
one is interested in counting the number of detections.  We refer to sites in $S_a$ as the destination
sites and sites in $S_v$ as the visitation sites. The sets $S_a$ and $S_v$ are mutually exclusive
and the set of defects is given by $S = S_a \cup S_v$ with the total number of defects being
$\lp S \rp = M_a + M_v = M$. Note that each site has its own independent probability of, respectively, detectability and absorption, that is we have $\mvec{\rho} = (\rho_1, \cdots, \rho_M)$ with
$\rho_{i}$ for $1\leq i\leq M_a$ being the partial absorption probability at site $\boldsymbol{r}_i$, while $\rho_{j}$ for $M_a+1\leq j\leq M$ being the detection probability of a visit at site $\boldsymbol{r}_j$. With this set-up, by generalising the formalism shown in \cite{rubinweiss1982}, it is possible to show that the generating function of the probability of (detected) visits at the sites of interest is given by
\begin{equation}
    \overline{V}_{\novec}(\mvec{x}, S, \mvec{\rho}) = 1 - \frac{\mdet{\mathcal{C}^{(d_1)} -\mathcal{C}^{(1)}} - \mdet{\mathcal{C}^{(d_2)} -\mathcal{C}^{(2)}}}{\mdet{\mathcal{C}^{(d_1)}} -\mdet{\mathcal{C}^{(d_2)}}},
    \label{eq:visits_x}
\end{equation}
where 
\begin{align}
  \label{cond:eq:cd1ij}
  \mathcal{C}^{(d_1)}_{i,j} &= 
  \lc
  \begin{matrix*}[l]
    \rho_{j} \mfptt{\rss[i]}{\rss[j]} (1 - \delta_{i, j})  + (\rho_{j} - 1) \mrett{\rss[i]} \delta_{i,j} ,
    &\ \   1 \leq j \leq M_a, \\
    \rho_{j} (1 - x_j)\mfptt{\rss[i]}{\rss[j]} (1 - \delta_{i, j})  + \ls \rho_{j} (1 - x_j) - 1 \rs \mrett{\rss[i]} \delta_{i,j} ,
    &\ \   M_a + 1 \leq j \leq M,
  \end{matrix*}
  \rd \\ 
  \label{cond:eq:cd2ij}
  \mathcal{C}^{(d_2)}_{i,j} &= 
    \mathcal{C}^{(d_1)}_{i,j} - 
  \lc
  \begin{matrix*}[l]
    \rho_{j},
    &\ \   1 \leq j \leq M_a, \\
    \rho_{j}( 1- x_j),
    &\ \   M_a + 1 \leq j \leq M ,
  \end{matrix*}
  \rd \\ 
  \label{cond:eq:c1ij}
  \mathcal{C}^{(1)}_{i,j} &= 
  \lc
  \begin{matrix*}[l]
    \rho_{j} \mfptt{\rss[i]}{\novec}, &\ \   1 \leq j \leq M_a, \\
      0, &\ \   M_a +  1 \leq j \leq M,
  \end{matrix*}
  \rd \\ 
  \label{cond:eq:c2ij}
    \mathcal{C}^{(2)}_{i,j} &= 
  \lc
  \begin{matrix*}[l]
    \rho_{j} (\mfptt{\rss[i]}{\novec} - 1) &\ \   1 \leq j \leq M_a, \\
      0, &\ \   M_a + 1 \leq j \leq M,
  \end{matrix*}
  \rd
\end{align}
with the notation $\overline{V}_{\novec}(\mvec{x}, S, \mvec{\rho}) = \sum_{k_1} \cdots \sum_{k_{M_v}}V_{\novec}(k_1, \cdots, k_{M_v}, S, \mvec{\rho}) x^{k_1} \cdots x^{k_{M_v}}$ representing the multi-dimensional generating function over the visits $k_1, \cdots, k_{M_v}$ to the sites $\rss[1], \cdots, \rss[M_v]$.
\begin{figure}
\centering
\includegraphics[width=\textwidth]{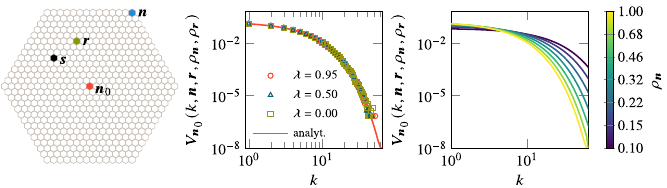}
\caption{The probability of the number of visits at site $\rss[]{}$ before being absorbed at $\nvec$ in the presence of a sticky site $\mvec{s}$. Left most panel shows a schematic of the spatial arrangement. We use a periodic hexagonal propagator of circumradius $R=13$, yielding a total of 547 sites, a LRW diffusivity of $q = 1$, and perfect detection at site $\boldsymbol{r}$, i.e. $\rho_{\mvec{r}} = 1.0$. The sticky heterogeneity is constructed such that the movement probability at the site is $q_{\mvec{s}} = q (1 - \lambda) $. Centre panel, probability of $k$ number of visits, when $\rho_{\mvec{n}} = 1.0$, for different values of $\lambda$. The marks, indicating $10^6$ simulations, all collapse onto the the analytic curve obtained by numerically inverting \cref{eq:visits_x}. Right most panel shows how the probability of $k$ visits depends on the reaction probability at the target site, i.e. $\rho_{\nvec}$.}
\label{fig:visit_comp}
\end{figure}

From Eqs. (\ref{cond:eq:cd1ij}-\ref{cond:eq:c2ij}) it is easy to realise that   $\overline{V}_{\novec}(\mvec{x}, S, \mvec{\rho})$ depends upon mean first-passage times and mean return times between the initial condition and the defective sites as well as between the defective sites, both the lattice of interest (visitation sites) and the absorbing lattice sites (terminating sites). When the lattice contains inert spatial heterogeneities, the disorder is expected to affect $\overline{V}_{\novec}(\mvec{x}, S, \mvec{\rho})$ through the mean first-passage and mean return times dependence on the locations and strengths of such heterogeneities in Eqs. (\ref{cond:eq:cd1ij}-\ref{cond:eq:c2ij}).

It turns out that when the disorder is due to sticky sites, $\overline{V}_{\novec}(\mvec{x}, S, \mvec{\rho})$ as well as $V_{\novec}(\mvec{k}, S, \mvec{\rho})$, that is the probability of the number of $k$ visits before being absorbed at any of the reactive sites, is independent of the numbers, strength and location of such heterogeneities. We show this indifference phenomenon in the simpler scenario of one sticky site at coordinate $\boldsymbol{s}$, one visitation site at $\boldsymbol{r}$ and one absorbing site at $\boldsymbol{n}$ in a periodic hexagonal domain. In Fig. \ref{fig:visit_comp} we plot $V_{\boldsymbol{n}_0}(k, \boldsymbol{n},\boldsymbol{r},\rho_{\boldsymbol{n}},\rho_{\boldsymbol{r}})$, that is the probability of $k$ visits before being absorbed at $\boldsymbol{r}$ as a function of $k$ for different choice of the heterogeneous parameter $\lambda$ and absorption probability $\rho_{\boldsymbol{n}}$ and detection probability $\rho_{\boldsymbol{r}}$.
The left most panel of Fig. \ref{fig:visit_comp} shows the geometrical arrangement of the problem.
 In the centre most panel we show that for different properties of the sticky site, all of the simulations collapse onto the analytic prediction, while the right most panel shows how the $k$-dependent probability changes as one reduces the reaction probability $\rho_{\nvec}$ of the target site $\nvec$. This latter dependence is expected since a decrease in  $\rho_{\nvec}$ maintains the walkers in the system for a longer period of time, which in turn allows for increased number visits to the site $\mvec{r}$. 

While surprising at first,  the disorder indifference phenomenon shown in the central panel of Fig. \ref{fig:visit_comp}, has an intuitive explanation. It is the result of the 
isotropic nature of the sticky heterogeneity,
which slows down the walker in a symmetric fashion, that is in an equal manner irrespective of the direction with which it moves away from the heterogeneous site. In other words, the sticky site acts to slow down all trajectories in all directions, lengthening their time before being absorbed even though their actual spatial path remains unchanged. Hence, the fraction of trajectories that visit $\mvec{r}$ $k$-times before being absorbed at $\mvec{n}$ also remains unchanged.

Given that the distribution is independent of the sticky heterogeneity, it is expected that the mean is also independent of the heterogeneity, i.e.
    \begin{equation}
      \begin{aligned}
      \mathcal{V}_{\novec}(\nvec, \rss[], \rho_{\nvec}, \rho_{\rss[]}) 
      &= 
      \frac{\rho_{\rss[]}}{\color{black}\mrett{\rss[]}}
      \ls
      {\color{black}
      \mfptt{\nvec}{\novec} + 
      \mfptt{\rss[]}{\nvec} - 
      \mfptt{\rss[]}{\novec}
      + \mrett{\nvec}} \lb \frac{1}{\rho_{\nvec}} - 1 \rb
      \rs. 
      \end{aligned}
      \label{eq:visit_mean}
    \end{equation}
Using the exact expressions for the mean first-passage time $T_{\boldsymbol{u}\rightarrow \boldsymbol{v}}$ and the mean return time $\mrett{\nvec}$ when sticky heterogeneities are present (see Eqs. (8) and (9) in ref. \cite{sarvaharmangiuggioli2023}) one can show that indeed $\mathcal{V}_{\novec}(\nvec, \rss[], \rho_{\nvec}, \rho_{\rss[]})$ is independent of the location and strength of the isotropic sticky site.

\section{Discussions, open problems and future directions}
\label{sec:concl}

While we have included a few new findings, we have focused on presenting a series of  quantitative tools that have recently appeared in the LRW literature and that we believe could be exploited in many scenarios both to understand empirical observations   and to make theoretical predictions.

In relation to the former, note that the challenge in defining the appropriate jump probabilities of the transition matrix is not different from the one faced by using models with continuous variables when extracting information about diffusion constants or other transport parameters. When the dynamics is relatively complex, e.g. in the presence of spatial disorder, a convenient approach is often that of comparing the dynamics observed in the absence of the underlying heterogeneities to theoretically obtained outputs, one prototypical example being the MSD in an unbounded domain from which the system diffusivity can be extracted. The analysis of the change in dynamics when the disorder is present then allows to identify the elements of the transition matrix  relative to the homogeneous environment. 

In regards to the theoretical advantages note that there is a computational gain in employing the analytical exact formulation as compared to a numerical iteration of the Master equation \cite{condaminetal2005}, e.g. in finding the first-passage probability and the mean first passage time to a single target in a hypercubic lattice of $d$ dimensions and length $N$ there is a factor of $N^{d}$ reduction in the time complexity \cite{giuggiolisarvaharman2022}.  More generally in the context of spatially disordered lattices, the advantage of our framework rests on being able to compute determinants of size equal to the number of spatial defects versus the numerical evaluation of eigenvalues and eigenvectors of a large sparse matrix whose size corresponds to the number of lattice sites, an approach suggested in ref. \cite{grebenkovtupikina2018}. 

Many areas in probability and combinatorics could benefit from these advances. For example the analysis of mixing time \cite{levinperes2017} and cut-off phenomena \cite{diaconis2011} in finite domains may find convenient to use the analytical expressions of the propagators and provide more rigorous bounds on known time scales.

The generality of the approach to study the dynamics on disordered lattice could also be exploited to study the dynamics of LRWs on networks \cite{masudaetal2017}. By adding, cutting or modifying the link strengths of regular lattices, we may build an arbitrary network. Starting from a lattice of which a propagator generating function is known, it would thus be possible to determine analytically the LRW propagator on the network by employing the formalism for inert spatial heterogeneities of Sec. \ref{sec:inert}.

There exist classical problems with Brownian walks that account for diffusion in some representative force field, e.g. a quadratic potential or the so-called V-potential. With the exception of ref. \cite{kac1947a} there has not been many attempts in the LRW literature to study these types of problems, but it would be beneficial to add them to the arsenal of tools in LRW theory.

While our theory can be easily applied to study interacting LRWs for dilute systems, in light of the already existing link between many-particle stochastic systems and quantum many-body dynamics, there are potential avenues to look  at some fundamental collective phenomena with some new techniques. One example is the classic asymmetric exclusion process on a 1D line \cite{stinchcombe2001} for which 
the exclusion interaction can be formulated by inserting reflecting barriers in the higher dimensional space spanned by the walkers.

We have discussed so far Markov problems, but there are plenty of movement processes for which the non-Markov nature of the process cannot be neglected~\cite{vilk2022ergodicity, bovet1988spatial, kenkre1983coherence}. While the amount of Markov literature would dwarf the non-Markov one if compared, there is great interest in developing a formalism for non-Markov LRWs. The majority of the work on the latter has been computational, while advances on the analytical front has been scarce, but see refs. \cite{schutztrimper2004,boyersolis-salas2014,falcon-cortesetal2017,boyeretal2019,meyerrieger2021} for some exceptions.

For the one-step non-Markovian LRW, the correlated LRW, analytic results have appeared in the past for the full first-passage probability in unbounded domains \cite{larralde2020first} and for periodically bounded domains only in terms of the mean first-passage time \cite{tejedor2012optimizing}. Recently, a general framework has been developed \cite{marrisgiuggioli2023}, which allows for the exact determination of the full first-passage probability of a bounded correlated LRW on both the hypercubic and the hexagonal lattice. Subordination procedures may also be exploited to generate subdiffusive LRWs, along the lines of the one employed in ref. \cite{schulzetal2014}, whereby  a subordination of the number of steps of LRW by a continuous time renewal counting process has led to subdiffusive RWs.

We conclude by thanking the book editors for the opportunity to write about the topic and with the hope that this short review will help raise awareness of the strengths and possibilities when one study transport processes using LRWs.

\section*{Acknowledgements}
LG acknowledges funding from the Biotechnology and Biological Sciences Research Council (BBSRC) Grant No. BB/T012196/1, and the National Environment Research Council (NERC) Grant No. NE/W00545X/1. All authors would like to thank the Isaac Newton Institute for Mathematical Sciences for support and hospitality during the programme `Mathematics of Movement: an interdisciplinary approach to mutual challenges in animal ecology and cell biology', when the work on this chapter was undertaken, supported by the EPSRC Grant Number EP/R014604/1. All authors wish to thank the book editors, Denis Grebenkov, Ralf Metzler, and Gleb Oshanin, for the invitation to contribute.

\end{document}